\documentclass[notoc]{JHEP3}
\usepackage{graphicx,psfrag}
\usepackage{color,cite}
\let\normalcolor\relax

\newcommand{\eqref}[1]{(\ref{#1})}
\newcommand{\text}[1]{{\rm #1}}

\newenvironment{changemargin}[2]{%
  \begin{list}{}{%
    \setlength{\topsep}{0pt}%
    \setlength{\leftmargin}{#1}%
    \setlength{\rightmargin}{#2}%
    \setlength{\listparindent}{\parindent}%
    \setlength{\itemindent}{\parindent}%
    \setlength{\parsep}{\parskip}%
  }%
  \item[]}{\end{list}}

\newcommand{\be}{\begin{equation}}
\newcommand{\ee}{\end{equation}}
\newcommand{\cmb}{\begin{changemargin}}
\newcommand{\cme}{\end{changemargin}}
\newcommand{\bea}{\begin{eqnarray}}
\newcommand{\eea}{\end{eqnarray}}
\def\spa#1.#2{\langle#1\,#2\rangle}
\def\spb#1.#2{[#1\,#2]}
\def\sandmm#1.#2.#3{%
\left\langle\smash{#1}{\rphantom1}\right|{#2}%
\left|\smash{#3}{\rphantom1}\right]}
\def\spab#1.#2.#3{\sandmm#1.#2.#3}
\def\spba#1.#2.#3{\sandpp#1.#2.#3}
\def\spaa#1.#2.#3.#4{\sandmp#1.{#2#3}.#4}
\def\spbb#1.#2.#3.#4{\sandpm#1.{#2#3}.#4}
\def\spash#1.#2{\spa{\smash{#1}}.{\smash{#2}}}
\def\spbsh#1.#2{\spb{\smash{#1}}.{\smash{#2}}}

\def\ksl{\not{\hbox{\kern-2.3pt $k$}}}

\title{Jet Shape Resummation Using Soft-Collinear Effective Theory}

\author{Yang-Ting Chien and Ivan Vitev\\
Theoretical Division, T-2\\
Los Alamos National Laboratory\\
Los Alamos, NM 87545, USA}

\abstract{
The jet shape is a classic jet substructure observable that probes the average transverse energy profile inside a reconstructed jet.
The studies of jet shapes in proton-proton collisions have served as precision tests of perturbative Quantum Chromodynamics (QCD). They have also recently become the baseline for studying the in-medium modification of parton showers in ultra-relativistic nucleus-nucleus collisions.
The jet shape is a function of two angular parameters $R$ and $r$, which can be at hierarchical scales. Its calculation suffers from
large logarithms of the ratio between the two scales, and these phase space logarithms can be conveniently  resummed in the framework of soft-collinear effective theory (SCET). We find that, up to power corrections, the integral jet shape can be expressed in a factorized form which involves only the ratio between two jet energy functions. Resummation is performed at next-to-leading logarithmic order using renormalization-group evolution techniques. Comparisons to jet shape measurements at the Large Hadron Collider (LHC) are presented to verify the dominant role of the collinear parton shower and to identify the kinematic region in which power-suppressed soft modes and non-perturbative effects may play a role.
}

\begin{document}
\section{Introduction}
\label{sec:intro}
Studies of jet substructure provide precision tests of perturbative QCD in high energy processes. They originate from the studies of event shapes in $e^+e^-$ collisions, which helped test and confirm the gauge theory structure of QCD~\cite{Farhi:1977sg, Georgi:1977sf, PhysRevLett.41.1581, PhysRevLett.41.1585, PhysRevD.19.2018, Heister:2003aj, Abdallah:2003xz, Achard:2004sv, Abbiendi:2004qz}. Recently, accurate event shape calculations also allowed one of the most precise extraction of the strong coupling constant~\cite{Becher:2008cf, Chien:2010kc, Davison:2008vx, Abbate:2010xh}. However, at hadron colliders, due to the presence of beam remnants, underlying event and pileup, the studies of event-wide inclusive observables become much more complicated. Instead, investigation of exclusive jet substructure observables attracts more attention, and considerable progress has been made in this direction \cite{Altheimer:2013yza}. One of the goal of such studies is to help distinguish possible signals of new physics beyond the Standard Model from large QCD backgrounds. An important problem, for example, is to develop improved methods to distinguish quark-initiated from gluon-initiated jets \cite{Gallicchio:2011xq,Gallicchio:2012ez}. Advances in this area will have numerous applications in new physics searches.

\begin{figure}[t]
    \begin{center}
    \psfrag{x}{$r$}
    \psfrag{y}{$\Psi(r)$}
    \psfrag{z}{$\frac{d\Psi(r)}{dr}$}
    \includegraphics[scale=0.95, trim = 0mm 0mm 0mm 0mm , clip=true]{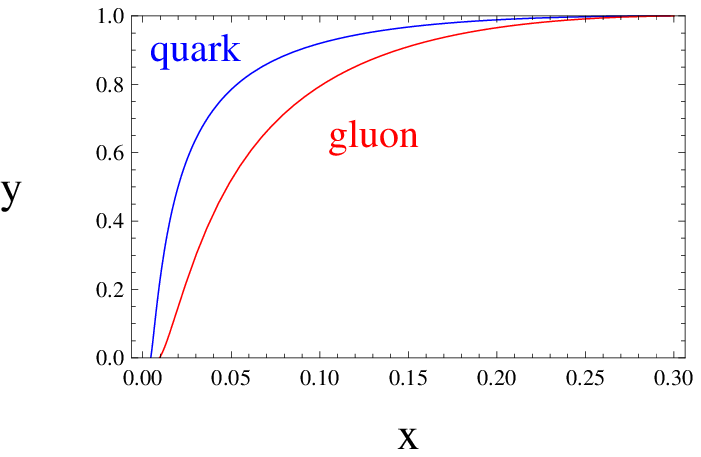}~~~
    \includegraphics[scale=0.95, trim = 0mm 0mm 0mm 0mm , clip=true]{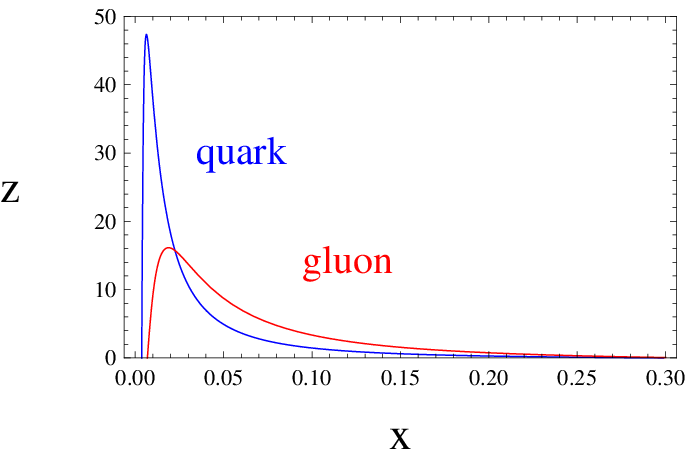}
    \caption{\label{fig:JS} The integral (left) and differential (right) jet shapes of {\color{blue}{quark}} and {\color{red}{gluon}} jets of size $R=0.3$ in proton-proton collisions, plotted as an illustration of their differences. Jet shape contains the information about the transverse energy distribution inside a jet. On average, quark jets are more localized whereas gluon jets are more spread out.}
    \end{center}
\end{figure}

Among the observables instrumental in quark-gluon discrimination, a classic jet substructure observable called the jet shape~\cite{Ellis:1992qq} has been studied for more than two decades. The integral jet shape is the average fraction of the transverse energy of the jet measured within a subcone of size $r$, smaller than the size $R$ of the jet, around the jet axis which is conventionally chosen to be along the 3-momentum of the jet. The differential jet shape is then the derivative with respect to $r$. The jet shape probes the transverse energy distribution inside a jet, which is very different for quark-initiated and gluon-initiated jets. Typically, quark jets are more localized whereas gluon jets are more spread out. This can be seen from the locations of the peaks of the differential jet shape distributions in Figure~\ref{fig:JS}, shown here as an illustration. Historically, the jet shape was introduced and calculated in QCD at leading-order in~\cite{Ellis:1992qq}. The observable was later resummed using the modified leading logarithmic approximation~\cite{Seymour:1997kj}. The contributions from initial state radiation and non-perturbative effects were also examined. A phenomenological parameter $R_{\rm sep}$~\cite{Ellis:1992qq, Seymour:1997kj} can be used to fit the data with the leading-order results. For more precise comparison with the experimental data a next-to-leading order calculation is required, but the result is not available at this time. Refs.~\cite{Li:2011hy, Li:2012bw} give another resummation framework using perturbative QCD and a comparison with the Tevatron and the LHC data.

On the other hand, the studies of jet shapes in heavy ion collisions have drawn considerable attention in the high-energy nuclear physics community. One of the top priorities of the heavy ion program at the LHC is to determine the properties of the hot, dense medium which is produced in ultra-relativistic nuclear collisions and referred to as the quark-gluon plasma (QGP). In such highly energetic collisions of ions, jets are produced and subsequently quenched as they propagate through the medium~\cite{Vitev:2008rz,He:2011pd}. The jet quenching phenomena give strong evidence for the creation of the QGP~\cite{Aad:2012vca,Aad:2010bu,Chatrchyan:2011sx} and build upon the well-established leading particle suppression pattern. The modification of jet shapes provides unique information about the structure of the in-medium parton showers. The first  measurement of the modification of jet shapes in lead-lead (Pb+Pb) collisions with small experimental uncertainties was performed by the CMS collaboration~\cite{Chatrchyan:2013kwa,Kurt:2014mca}. On the theory side, Ref.~\cite{Vitev:2008rz} builds upon the jet shape calculations in proton-proton collisions~\cite{Seymour:1997kj} and studies the medium modification of jet shapes using the Gyulassy-Levai-Vitev formalism~\cite{Gyulassy:2000er,Gyulassy:2000fs} in the soft gluon limit. Monte Carlo transport simulations of jet shapes in Pb+Pb collisions at the LHC have also been recently performed~\cite{Ma:2013uqa,Ramos:2014mba}. However, a study going beyond these approximations and addressing the precision of the jet shape calculations in a systematically improvable way is needed in both proton-proton and heavy ion collisions.

In this paper, we focus on the jet shape calculations in proton-proton collisions using soft-collinear effective theory (SCET)~\cite{Bauer:2000ew, Bauer:2000yr, Bauer:2001ct, Bauer:2001yt, Bauer:2002nz}. The calculations in heavy ion collisions using SCET with Glauber gluon interactions in the medium \cite{Idilbi:2008vm, Ovanesyan:2011xy}, and the full medium-induced splitting functions~\cite{Ovanesyan:2011kn,Fickinger:2013xwa} and
applications~\cite{Kang:2014xsa} will be discussed in a forthcoming paper~\cite{chien:2014jsm}. SCET is an effective field theory of QCD for processes with energetic light-like and soft degrees of freedom with a systematic power counting. In events with highly collimated jets, the power counting parameter $\lambda\sim m_J/Q$, which is the ratio between the jet mass and the jet energy, is small and the leading power contribution calculated in SCET is a very good approximation of the full QCD result. SCET separates physics at different energy scales, and the factorization of the hard, collinear and soft sectors is more transparent. The hard, jet and soft functions involved in the factorization theorem of a physical cross section, as well as their anomalous dimensions, can be calculated order by order at each characteristic scale. Large logarithms of the ratio between hierarchical energy scales are resummed through the renormalization-group evolution of these functions.

In the context of jet shape calculations, as we will see below, for small radii $R$ the observable depends mostly on the jet energy and the partonic origin of the jet. It is not sensitive to the details of the underlying hard scattering processes as well as the soft radiation in the whole event.  The contribution from the soft radiation to the jet shape is power suppressed. Therefore, the collinear sector is dominant in such calculations and the factorized expression for the jet shape involves only the ratio between two jet energy functions, which we define and calculate at leading order (LO). We then identify two characteristic jet scales, each proportional to the angular scale ($r$ or $R$) within which we measure the transverse energy. The logarithms of the ratio between the two jet scales are exactly the logarithms of the form $\ln r/R$, which we can resum through the renormalization-group evolution of the jet energy functions between the two jet scales. With the two-loop cusp anomalous dimension, the two-loop running of the strong coupling constant and the one-loop anomalous dimension of the jet energy function, the jet shape is resummed to next-to-leading logarithmic (NLL) accuracy \footnote{ By $\rm N^{k}LL$ we mean the resummed series includes terms of the form $\alpha_s^n\ln^{m}r/R$ with $2n \geq m \geq 2n-2k+1$. We use this convention in the region of validity of $r$ where $\alpha_s \ln^2 r/R \lesssim 1$, which is the case in comparing with both the CMS and ATLAS measurements, as we will see. 
However, in the region where $\alpha_s \ln r/R \sim 1$, more terms should be resummed using the convention which is commonly referred to as counting in the exponent.
\cite{Almeida:2014uva} gives a useful
discussion about the counting of precisions in both perturbative QCD and SCET calculations. Note that non-global logarithms affect the $\alpha_s^2\ln^2 r/R$ terms at NNLL in the convention above. }. However, in this work we do not include the contributions from initial state radiation and non-perturbative effects. We ignore power suppressed terms of ${\cal O}(R)$ and focus on the resummation of large logarithms. Note that terms of ${\cal O}(r/R)$ can still be large at $r \approx R$ and they are properly captured by the SCET formalism. To systematically extend the precision to next-to-leading order (NLO) and next-to-next-to leading logarithmic (NNLL) accuracy, we will need the two-loop jet energy function and its two-loop anomalous dimension. At this order the issue about non-global logarithms \cite{Dasgupta:2001sh} and the way to resum them will also arise. We will leave these interesting topics for future work.

The rest of the paper is organized as follows. In section~\ref{sec:obs} we give the definition of the jet shape and discuss the choice of the jet axis, which is related to the form of the factorized expression. In section~\ref{sec:fac} we discuss the power counting and the calculation of the jet shape in SCET. We show that the contribution from soft particles to the jet shape is power suppressed if we choose a soft-recoil free
axis~\cite{Bertolini:2013iqa,Larkoski:2014uqa} in the jet shape definition. The factorized expression of the jet shape has a simple product form without recoil-momentum convolution. This leads to the cancellation of the hard and soft functions in the factorized expression for the jet shape, which involves only the ratio of the jet energy functions. We also give the operator definition of the jet energy function and calculate it at one-loop for both quark and gluon jets reconstructed using the cone or the anti-$\rm k_T$ algorithm \cite{Cacciari:2008gp}. The resummation of jet shapes is performed to NLL accuracy using the renormalization-group techniques, and we estimate the uncertainties of our calculations by varying the characteristic jet scales. In section \ref{sec:results} we compare our resummed jet shape results with the {\sc pythia} 8 simulations and the CMS measurements. We present our conclusion and give an outlook in section \ref{sec:conc}.

\section{The observable}
\label{sec:obs}

In this section we will give the jet shape definition and discuss some of the related subtleties. Before we can study any property of a jet, we need to precisely define what a jet is. This is conventionally done using a jet algorithm with a parameter $R$, which characterizes the size of the jet. Different jet algorithms will give jets with different substructures. For a jet reconstructed using a jet algorithm, we first define a jet axis $\hat n$. A natural choice of $\hat n$ is the direction of the 3-momentum of the jet.
However, such axis is not necessarily along the direction of the dominant energy flow within the jet. The factorized expression for an observable referencing this axis will involve an intricate convolution over the recoil momentum between the collinear and the soft sectors. On the other hand, there are choices of $\hat n$, e.g. the broadening axis or the winner-take-all axis \cite{Bertolini:2013iqa, Larkoski:2014uqa}, which are soft-recoil free.
These axes absorb the recoil sensitivity and point along the collinear momentum which gives a simpler factorized form without recoil-momentum convolution. We will come back to this point in more details in section \ref{sec:fac}.

Given a jet with an axis $\hat n$, its integral jet shape $\Psi_J(r)$ is defined as follows,
\be
    \Psi_J(r)=\frac{\sum_{i,~d_{i\hat n}<r} E^i_T}{\sum_{i,~d_{i\hat n}<R} E^i_T}\;,
\ee
which is the fraction of the transverse energy $E_T$ of the jet within an angular scale $r$ from the jet axis. The transverse energy is measured with respect to the beam direction. By definition $\Psi_J(R)=1$. Here, $d_{i\hat n}$ is the distance metric between the $i$-th particle in the jet and the jet axis $\hat n$. It can be the Euclidean distance between the two directions along the particle and the jet axis on the rapidity-azimuthal angle $(y,\phi)$ plane,
\be
    d_{i\hat n}=\sqrt{(y_i-y_{jet})^2+(\phi_i-\phi_{jet})^2}\;.
\ee
Note that, with this definition the jet shape is boost invariant along the beam direction. So we can calculate the jet shape in the frame where the jet is central ($y_{jet}=0$). In that frame, up to power corrections, the metric is equivalent to
\be
    d_{i\hat n}=\cos^{-1}({\hat n_i}\cdot{\hat n})\;,
\ee
which is the angle between the momentum of the $i$-th particle and the jet axis in 3-space. Also, up to power corrections we can consider the energies instead of the transverse energies of particles in the central jet. So, in this paper,
\be
    \Psi_J(r)=\frac{\sum_{i,~\theta_{i\hat n}<r} E^i}{\sum_{i,~\theta_{i\hat n}<R} E^i}\equiv\frac{E_r}{E_R}\;,
\ee
setting $y_{jet}=0$. In experiment, we measure the {\sl averaged} integral jet shape $\Psi(r)$ (we will drop the word {\sl averaged} and refer to $\Psi(r)$ as the integral jet shape from now on),
\be
    \Psi(r)=\frac{1}{N_J}\sum_{J=1}^{N_J} \Psi_J(r)\;.
\ee
The differential jet shape $\psi(r)$ is defined to be its derivative,
\be
    \psi(r)=\frac{d\Psi(r)}{dr}\;,
\ee
which tells us how the energy inside the jet is distributed in $r$.  Recall that the jet shape has dependence on the jet algorithm and the parameter $R$ used in the jet definition, which we suppress here for notational simplicity. At the Tevatron mostly the iterative cone algorithm was used, whereas at the LHC the anti-$\rm k_T$ algorithm is used almost exclusively.

The jet shape is a function of $r$ and $R$, which are parameters or angular coordinates. In the jet shape calculations terms of the form $\alpha_s^n\ln^m r/R$ appear, which can become large if $r\ll R$. In this regime the fixed order expansion breaks down and the large logarithms need to be resummed. The resummation was performed some time ago using the modified leading logarithmic approximation (MLLA) \cite{Seymour:1997kj}, including the contributions from initial state radiation and non-perturbative effects. A phenomenological parameter $R_{\rm sep}$, which can be thought of as the effective separation between the particles at leading order, can be used to fit with the experimental data. In this paper we will not follow this phenomenological approach.

In the next section we will calculate the jet shape using SCET. 
Large logarithms come from the presence of multiple, hierarchical energy scales in the problem. This is the situation in which effective field theory techniques are useful because the corrections to the leading power contribution are suppressed by a small power counting parameter. The key ingredient in this approach is the factorization, which separates physics into multiple, single-scaled sectors. Large logarithms are then resummed by the renormalization-group evolution of different pieces of physics between their natural scales.

\section{Factorized expression for the jet shape}
\label{sec:fac}

\begin{figure}[t]
    \begin{center}
    \includegraphics[scale=0.5, trim = 0mm 0mm 0mm 0mm , clip=true]{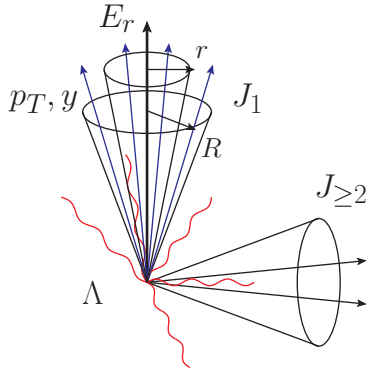}
    \caption{\label{fig:event}Schematic event topology of $N$-jet production with {\color{blue}{collinear}} and {\color{red}{soft}} radiation. Jets are reconstructed using a jet algorithm with a parameter $R$. The energy $E_r$ inside a cone of size $r$ in $J_1$ is measured, as well as its transverse momentum $p_T$ and rapidity $y$. An energy cutoff $\Lambda$ outside the jets is imposed to ensure the $N$-jet configuration.}
    \end{center}
\end{figure}

For concreteness, let us consider the shapes of jets from an $N$-jet configuration in $e^+e^-$ collisions without loss of generality, as we will see \footnote{The $N$ jets are assumed to be energetic by passing a hard $p_T$ cut so that the power counting and the factorized expression we will write down are valid. At hadron colliders, there are power-suppressed contributions of ${\cal O}(R)$ from initial state radiation which we will neglect.}. Jets are reconstructed using either the cone or the anti-$\rm k_T$ algorithm with a parameter $R$ that is parametrically small. At the LHC a typical $R=0.5$ to 0.7 is chosen in the physics analyses of proton-proton collisions. For heavy ion collisions, because of the underlying event contaminations a even smaller $R=0.3$ is chosen at CMS. In these cases jets are highly collimated and the process can be accurately described by the soft-collinear effective theory (SCET). An energy cutoff $\Lambda$ outside the jets is imposed to ensure an $N$-jet configuration. We also measure the energy $E_r$ inside a cone of size $r$ in one jet (labeled by 1), as well as the transverse momenta ${p_T}_i$ and pseudo-rapidity $y_i$ of all the jets.

Before we write down the factorized expression for the jet shape in SCET, let us study the power counting of the observable. In light-cone coordinates with $p=(\bar n\cdot p, n\cdot p, \vec p_\perp)$ where $n=(1,\hat n)$ and $\bar n=(1,-\hat n)$, the momentum scalings of the collinear and the soft 
particles are
\be
    p_c=Q(1,\lambda^2,\lambda)\;,~p_s=Q(\lambda,\lambda,\lambda)\;,
\ee
where $Q$ is the center of mass energy of the $e^+e^-$ collisions and $\lambda$ is the power counting parameter which describes how wide a jet is spread out. There is one collinear sector for for each jet with the collinear direction $n$. The power counting parameter satisfies $R\gtrsim \lambda$ so that most of the jet energy is included in the jet reconstruction. The jet energy $E$ has contributions $E^c$ and $E^s$ from both the collinear and the soft 
sectors,
\be
    \Psi_J(r)=\frac{E_r}{E_R}=\frac{E_r^c+E_r^s}{E_R^c+E_R^s}=\frac{E_r^c}{E_R^c}+{\cal O}(\lambda)\;.
\ee
Up to power corrections, the integral jet shape can be calculated using only the collinear momenta, and the soft contributions can be neglected.
This approximation works best  when $R$ is not large. On the other hand, as was briefly discussed in section \ref{sec:obs}, soft radiation can potentially alter the collinear momentum direction by an ${\cal O}(\lambda)$ amount. For $r$ of ${\cal O}(\lambda)$ or smaller, soft recoil can actually change $E_r$ by an ${\cal O}(1)$ amount which will spoil the simple factorized expression. This can be remedied by choosing the recoil-free
jet axes for the jet shape measurements \footnote{This has not been implemented in experimental measurements to-date, and it introduces another power correction when comparing our calculations with experiments.}. Such jet axes will always point along the collinear momentum directions \cite{Bertolini:2013iqa, Larkoski:2014uqa}. To avoid the issue of recoil from the soft particles outside the jets, we impose a constraint on the energy cutoff $\Lambda$ by demanding $\Lambda/Q\ll R$. At hadron colliders, because of dynamical threshold enhancement the partonic phase space where the jets have small jet masses dominates in the cross section calculations \cite{Appell:1988ie, Catani:1998tm, Becher:2007ty, Chien:2012ur}. The cross section of jet configurations with large jet masses is suppressed because parton distribution functions die off quickly when the momentum fraction of the parton is close to one. This will reduce the recoil sensitivity of the jet axis and therefore it is less of an issue.

Detailed derivations and discussions about the factorized expression similar to the one we will write down here can be found in \cite{Bauer:2003di,Ellis:2010rwa, Larkoski:2014uqa}. The differential cross section for $N$-jet production with jets $p_{T_i}$ and $y_i$, an energy $E_r$ inside the cone of size $r$ in one jet (labeled by 1), and an energy cutoff $\Lambda$ outside all the jets is the following \footnote{The functional dependence on $r$ and $R$ are suppressed for notational simplicity.},
\bea
    \frac{1}{\sigma_0}\frac{d\sigma}{dE_rdp_{T_i}dy_i}&=&H(p_{T_i},y_i,\mu)J_{\omega_1}(E_r,\mu) J_{\omega_2}(\mu)\dots J_{\omega_N}(\mu)S_{n_1n_2\dots n_N}(\Lambda,\mu)\nonumber  \\
    &&+{\cal O}\Big(\frac{\Lambda}{Q}\Big)+{\cal O}(R)\;.
\eea
Here, $H(p_{T_i},y_i,\mu)$ is the hard function which contains the information about the $N$-jet productions at high scale $Q$ and is independent of the jet shape measurements. It is the square of the Wilson coefficient when we match QCD and SCET at the hard scale. $J_{\omega}(E_r,\mu)$ is a newly defined jet function, which is the probability of measuring an energy $E_r$ inside a $r$ cone for a jet of size $R$ with $\omega=2E_J=2p_T\cosh y$,
\be
    J_{\omega}(E_r,\mu)=\sum_{X_c}\langle 0|\bar\chi_{\omega}(0)|X_c\rangle\langle X_c|\chi_{\omega}(0)|0\rangle\delta(E_r-\hat E^{<r}(X_c))\;.
\ee
Here, $\chi_\omega$ is the collinear jet field in SCET, and the operator $\hat E^{<r}$ returns the energy of the collinear particles $X_c$ inside the $r$ cone. By imposing an energy cutoff $\Lambda$ outside the jets, we are essentially restricting the collinear radiation to be all inside the jets up to corrections of ${\cal O}(\Lambda/Q)$. All the other jet functions without the jet energy measurements are the "unmeasured" jet functions \cite{Ellis:2010rwa}. Once we integrate out the collinear modes, we are left with a soft sector which is described by soft Wilson lines along the jet directions. The soft function is defined as follows,
\be
    S_{n_1n_2\dots n_N}(\Lambda,\mu)=\sum_{X_s}\langle 0|{\cal O}_s^\dagger(0)|X_s\rangle\langle X_s|{\cal O}_s(0)|0\rangle \Theta (\Lambda-\hat E^{>R}(X_s))\;,
\ee
where ${\cal O}_s(0)$ consists of $N$ soft Wilson lines along the $n_{1,2,\dots, N}$ directions intersecting at the origin 0. The operator $\hat E^{>R}$ returns the energy of the soft particles $X_s$ outside all  $N$ jets. Note that the factorized form is a simple product of the hard, jet and soft functions without any convolution. This is because the soft particles don't contribute to the jet energy at leading power and we choose a recoil-free
jet axis in the jet shape definition. All the jet and soft functions have $R$ dependence which we suppress for brevity.

Similarly, for the differential jet rate of $N$-jet production,
\be
    \frac{1}{\sigma_0}\frac{d\sigma}{dp_{T_i}dy_i}=H(p_{T_i},y_i,\mu)J_{\omega_1}(\mu) J_{\omega_2}(\mu)\dots J_{\omega_N}(\mu)S_{n_1n_2\dots n_N}(\Lambda,\mu)+{\cal O}\Big(\frac{\Lambda}{Q}\Big)+{\cal O}(R)\;,
\ee
with the {\sl same} hard, unmeasured jet (from 2 to $N$) and soft functions. The only difference is that we don't measure the energy of jet 1, so $J_{\omega_1}(E_r,\mu)$ is replaced by the unmeasured jet function $J_{\omega_1}(\mu)$. Now, the {\sl averaged} energy inside the cone of size $r$ in jet 1 with $\omega=\omega_1$ is
\bea
    \langle E_r\rangle_{\omega_1}
    &=&\frac{\int dE_r E_r\frac{1}{\sigma_0}\frac{d\sigma}{dE_rdp_{T_i}dy_i}}{\frac{1}{\sigma_0}\frac{d\sigma}{dp_{T_i}dy_i}}
    =\frac{H(p_{T_i},y_i,\mu)J^{E_r}_{\omega_1}(\mu) J_{\omega_2}(\mu)\dots J_{\omega_N}(\mu)S_{n_1n_2\dots n_N}(\Lambda,\mu)}{H(p_{T_i},y_i,\mu)J_{\omega_1}(\mu) J_{\omega_2}(\mu)\dots J_{\omega_N}(\mu)S_{n_1n_2\dots n_N}(\Lambda,\mu)}\nonumber\\
    &=&\frac{J^{E_r}_{\omega_1}(\mu)}{J_{\omega_1}(\mu)}\;.
\eea
Here,
\be
    J^{E_r}_{\omega}(\mu)=\int dE_r E_r~J_\omega(E_r,\mu)
\ee
is referred to as the jet energy function. Note that all the hard, unmeasured jet, and soft functions cancel in the calculations because of the product form of the factorized expression. The integral jet shape needs another average over the jet production cross sections, with proper phase space ($PS$) cuts on $p_T$ and $y$, and is therefore
\be
    \Psi(r)=\frac{1}{\sigma_{\rm total}}\sum_{i=q,g}\int_{PS} dp_Tdy \frac{d\sigma^{i}}{dp_Tdy}\Psi^i_\omega(r)\;,
\ee
where
\be
    \Psi_\omega(r)
    =\frac{\langle E_r\rangle_{\omega}}{\langle E_R\rangle_{\omega}}
    =\frac{J^{E_r}_{\omega}(\mu)/J_{\omega}(\mu)}{J^{E_R}_{\omega}(\mu)/J_{\omega}(\mu)}
    =\frac{J^{E_r}_{\omega}(\mu)}{J^{E_R}_{\omega}(\mu)}\;,
\ee
which is the ratio between two jet energy functions, and $\omega=2p_T$ in the frame where the jet is central ($y=0$). As we can see, the jet shape does not depend on the hard and soft functions nor on the information about the other jets. In other words, it is insensitive to the hard process and the soft radiation and depends only on the partonic origin and the energy of the jet.

From the factorized form  we can already infer some non-trivial properties of the jet and soft functions. By the renormalization-group invariance of the physical cross sections, the anomalous dimension of the soft function should be independent of the energy cutoff $\Lambda$ to all orders in perturbation theory. Therefore the soft anomalous dimension can only be a function of $R$ and is independent of the renormalization scale $\mu$. Also, the anomalous dimension of the jet energy function $J^{E_r}_\omega(\mu)$ is the same as the anomalous dimension of the unmeasured jet function $J_\omega(\mu)$ which does not depend on $r$. This makes $\Psi_\omega(r)$ renormalization-group invariant. Furthermore, the $R$ dependence in the anomalous dimensions of the jet and soft functions has to cancel because the hard function does not depend on $R$.

\subsection{One-loop jet energy functions}
\label{subsec:jetfn}
Having set up the factorization framework, now we proceed to calculate the jet energy function $J^{E_r}_{\omega}(\mu)$ at ${\cal O}(\alpha_s)$ for both quark jets and gluon jets reconstructed using the cone or anti-$\rm k_T$ algorithm. At this order, the initial collinear particle with momentum $l$ splits into two collinear particles with momenta $q$ and $l-q$. We use lightcone coordinates throughout the calculations where $k^+\equiv n\cdot k$ and $k^-\equiv\bar n\cdot k$. Also, we use dimensional regularization to regulate the divergences with the spacetime dimension $d=4-2\epsilon$, and the $\rm \overline{MS}$ renormalization scheme. For quark jets,
\bea
    J^{qE_r}_{\omega,{\rm alg}}(\mu)
    &=&4\pi\alpha_s\Big(\frac{\mu^2e^{\gamma_E}}{4\pi}\Big)^{\epsilon}C_F\int\frac{dl^+}{2\pi}\frac{1}{(l^+)^2}\int\frac{d^dq}{(2\pi)^d}
    \left[4\frac{l^+}{q^-}+(d-2)\frac{l^+-q^+}{\omega-q^-}\right]\nonumber\\
    &&\times 2\pi\delta(q^-q^+-q^2_\perp)\Theta(q^-)\Theta(q^+)~2\pi\delta\left(l^+-q^+-\frac{q^2_\perp}{\omega-q^-}\right)\Theta(\omega-q^-)\Theta(l^+-q^+)\nonumber\\
    &&\times\Big({\cal M}_1+{\cal M}_2+{\cal M}_3+{\cal M}_4\Big)\;,
\eea
and for gluon jets,
\bea
    J^{gE_r}_{\omega,{\rm alg}}(\mu)
    &=&8\pi\alpha_s\Big(\frac{\mu^2e^{\gamma_E}}{4\pi}\Big)^{\epsilon}\int\frac{dl^+}{2\pi}\frac{1}{l^+}\int\frac{d^dq}{(2\pi)^d}\left[T_Fn_f\left(1-\frac{2}{1-\epsilon}\frac{q^+q^-}{\omega l^+}\right) \right. \nonumber \\
 &&    \left. -C_A \left( 2-\frac{\omega}{q^-}-\frac{\omega}{\omega-q^-}-\frac{q^+q^-}{\omega l^+}\right)\right] 2\pi\delta(q^2)\Theta(q^-)\Theta(q^+)
 \nonumber\\
    && \times 2\pi\delta((l-q)^2)\Theta(\omega-q^-)\Theta(l^+-q^+)\times\Big({\cal M}_1+{\cal M}_2+{\cal M}_3+{\cal M}_4\Big)\;.
\eea
Here
\bea
    {\cal M}_1&=&\Theta\Big(\tan^2\frac{r}{2}-\frac{q^+}{q^-}\Big)\Theta\Big(\tan^2\frac{r}{2}-\frac{l^+-q^+}{\omega-q^-}\Big)\Theta_{\rm alg}\times l^0\\
    {\cal M}_2&=&\Theta\Big(\tan^2\frac{r}{2}-\frac{q^+}{q^-}\Big)\Theta\Big(\frac{l^+-q^+}{\omega-q^-}-\tan^2\frac{r}{2}\Big)\Theta_{\rm alg}\times q^0\\
    {\cal M}_3&=&\Theta\Big(\frac{q^+}{q^-}-\tan^2\frac{r}{2}\Big)\Theta\Big(\tan^2\frac{r}{2}-\frac{l^+-q^+}{\omega-q^-}\Big)\Theta_{\rm alg}\times (l^0-q^0)\\
    {\cal M}_4&=&\Theta\Big(\frac{q^+}{q^-}-\tan^2\frac{r}{2}\Big)\Theta\Big(\frac{l^+-q^+}{\omega-q^-}-\tan^2\frac{r}{2}\Big)\Theta_{\rm alg}\times 0\;,
\eea
which are the cases where each of the two particles are either inside the cone of size $r$ or not. The algorithm dependence enters in the calculations through $\Theta_{\rm alg}$ in the following way,
\bea
    \Theta_{\rm cone}
    &=&\Theta\left(\tan^2\frac{R}{2}-\frac{q^+}{q^-}\right)\Theta\left(\tan^2\frac{R}{2}-\frac{l^+-q^+}{\omega-q^-}\right)\nonumber\\
    \Theta_{\rm k_T}
    &=&\Theta\left(\tan^2\frac{R}{2}-\frac{q^+\omega^2}{q^-(\omega-q^-)^2}\right).
\eea
Note that at this order the phase space constraint is the same for anti-$\rm k_T$ jets and $\rm k_T$ jets, so we label the anti-$\rm k_T$ jet energy functions by $\rm k_T$ for simplicity. For cone jets the two collinear particles are constrained to be inside a cone of size $R$, whereas for anti-$\rm k_T$ jets the angle between the two collinear particles has to be smaller than $R$. Therefore one expect that cone jets are more spread out than anti-$\rm k_T$ jets.

We expand the dimensionally regularized one-loop jet energy functions in series of $\epsilon$'s. The anomalous dimension can be extracted from the coefficient of the $1/\epsilon$ pole in the series, and the ${\cal O}(\epsilon^0)$ piece will be the renormalized jet energy function in the $\overline{\rm MS}$ scheme. The jet energy functions for cone jets at ${\cal O}(\alpha_s)$ are
\bea
    \frac{2}{\omega}J^{qE_r}_{\omega,{\rm cone}}(\mu)
    &=&\frac{\alpha_s C_F}{2\pi}\left[\frac{1}{2}\ln^2\frac{\omega^2\tan^2\frac{r}{2}}{\mu^2}
    -\frac{3}{2}\ln\frac{\omega^2\tan^2\frac{r}{2}}{\mu^2}-2\ln X\ln\frac{\omega^2\tan^2\frac{r}{2}}{\mu^2} \right. \nonumber\\
    &+&2-\frac{3\pi^2}{4}+4{\rm Li_2}\Big(\frac{X}{1+X}\Big)+3\ln(1+X)+\frac{3X}{1+X}+2\ln^2(1+X)\nonumber\\
    &+& \left. \Big(-5\ln(1+X)+\frac{5X+2X^2}{1+X}-2X^2\ln \frac{X}{1+X}\Big)\tan^2\frac{R}{2}
    \right]\;,
\eea
and
\bea
    \frac{2}{\omega}J^{gE_r}_{\omega,{\rm cone}}(\mu)
    &=&\frac{\alpha_s}{2\pi}\left[\frac{C_A}{2}\ln^2\frac{\omega^2\tan^2\frac{r}{2}}{\mu^2}
    -\Big(\frac{11}{6}C_A-\frac{2}{3}T_Fn_f\Big)\ln\frac{\omega^2\tan^2\frac{r}{2}}{\mu^2}-2C_A\ln X\ln\frac{\omega^2\tan^2\frac{r}{2}}{\mu^2}  \right.\nonumber\\
    &-&\Big(\frac{5\pi^2}{12}-2{\rm Li_2}\Big(\frac{X}{1+X}\Big)+2{\rm Li_2}\Big(\frac{1}{1+X}\Big)\Big)C_A\nonumber\\
    &+&\Big(\frac{11}{3}C_A-\frac{4}{3}T_F n_f+2C_A\ln X\Big)\log(1+X)\nonumber\\
    &+&\frac{-2(5+63X+81X^2+35X^3)T_Fn_f+(65+351X+477X^2+203X^3)C_A}{36(1+X)^3}\nonumber\\
    &-&\Big(\frac{2X(6+13X+9X^2)T_Fn_f-X(36+85X+63X^2+12X^3)C_A}{6(1+X)^3}\nonumber\\
    &&\left. +2C_A X^2\ln X-2(T_Fn_f-C_A(3-X^2))\ln (1+X)\Big)\tan^2\frac{R}{2}
    \right]\;,
\eea
where
\be
    X=\frac{\tan\frac{r}{2}}{\tan\frac{R}{2}}\approx\frac{r}{R}~~~{\rm for}~r,R\ll1\;,
\ee
is boost-invariant along the jet direction. In fact, under a Lorentz boost with rapidity $\beta$ along the jet direction, $\omega$ and $r$ transform in the following way,
\be
    \omega\rightarrow e^\beta \omega\;,~\tan\frac{r}{2}\rightarrow e^{-\beta}\tan\frac{r}{2}\;.
\ee
Therefore the combination, $\omega\tan\frac{r}{2}$, is also boost-invariant. For anti-$\rm k_T$ jets, we have
\bea
    \frac{2}{\omega}J^{qE_r}_{\omega,{\rm k_T}}(\mu)
    &=&\frac{\alpha_s C_F}{2\pi}\left[\frac{1}{2}\ln^2\frac{\omega^2\tan^2\frac{r}{2}}{\mu^2}
    -\frac{3}{2}\ln\frac{\omega^2\tan^2\frac{r}{2}}{\mu^2}-2\ln X\ln\frac{\omega^2\tan^2\frac{r}{2}}{\mu^2} \right. \nonumber\\
    &+&\left. 2-\frac{3\pi^2}{4}+6X-\frac{3}{2}X^2-\Big(\frac{1}{2}X^2-2X^3+\frac{3}{4}X^4+2X^2\ln X\Big)\tan^2\frac{R}{2}
    \right]\;,~~~~~~~~~
\eea
and
\bea
    \frac{2}{\omega}J^{gE_r}_{\omega,{\rm k_T}}(\mu)
    &=&\frac{\alpha_s}{2\pi}\left[\frac{C_A}{2}\ln^2\frac{\omega^2\tan^2\frac{r}{2}}{\mu^2}
    -\Big(\frac{11}{6}C_A-\frac{2}{3}T_Fn_f\Big)\ln\frac{\omega^2\tan^2\frac{r}{2}}{\mu^2}-2C_A\ln X\ln\frac{\omega^2\tan^2\frac{r}{2}}{\mu^2}\right. \nonumber\\
    &+&\Big(\frac{65}{36}-\frac{3\pi^2}{4}+8X-3X^2+\frac{8X^3}{9}-\frac{X^4}{4}\Big)C_A\nonumber\\
    &+&\Big(-\frac{5}{18}-4X+3X^2-\frac{16X^3}{9}+\frac{X^4}{2}\Big)T_Fn_f\nonumber\\
    &-&\frac{X^2}{30}\Big((25-80X+45X^2-16X^3+5X^4+60\ln X)C_A\nonumber\\
    &+&\left.(-20+40X-45X^2+32X^3-10X^4)T_Fn_f\Big)\tan^2\frac{R}{2}
    \right]\;.
\eea
An important observation is that the choice of the renormalization scale $\mu=\omega\tan\frac{r}{2}$ eliminates all the large logarithms at ${\cal O}(\alpha_s)$. Also, the terms from the second line down in each expression of the jet energy function are non-singular when $X\rightarrow0$. The jet energy functions for jets reconstructed using different algorithms differ by these non-singular terms at this order. For $r=R$, we have
\bea
    \frac{2}{\omega}J^{qE_R}_{\omega,{\rm cone}}(\mu)
    &=&J^q_{\omega,{\rm cone}}(\mu)+\frac{\alpha_s}{2\pi}C_F\left(\frac{7}{2}-3\ln2\right)\tan^2\frac{R}{2}\nonumber\\
    \frac{2}{\omega}J^{gE_R}_{\omega,{\rm cone}}(\mu)
    &=&J^g_{\omega,{\rm cone}}(\mu)+\frac{\alpha_s}{2\pi}\left[C_A\Big(\frac{49}{12}-4\ln2\Big)-T_Fn_f\Big(\frac{7}{6}-2\ln2\Big)\right]\tan^2\frac{R}{2}\;,~~~~
\eea
and
\bea
    \frac{2}{\omega}J^{qE_R}_{\omega,{\rm k_T}}(\mu)
    &=&J^q_{\omega,{\rm k_T}}(\mu)+\frac{\alpha_s}{2\pi}\frac{3}{4}C_F\tan^2\frac{R}{2}\nonumber\\
    \frac{2}{\omega}J^{gE_R}_{\omega,{\rm k_T}}(\mu)
    &=&J^g_{\omega,{\rm k_T}}(\mu)+\frac{\alpha_s}{2\pi}\left[\frac{7}{10}C_A+\frac{1}{10}T_Fn_f\right]\tan^2\frac{R}{2}\;.
\eea
Here
\bea
    J^q_{\omega,{\rm alg}}(\mu)&=&\frac{\alpha_s}{2\pi}\left(-\frac{3}{2}C_F\ln\frac{\omega^2\tan^2\frac{R}{2}}{\mu^2}+\frac{1}{2}C_F\ln^2\frac{\omega^2\tan^2\frac{R}{2}}{\mu^2}
    +d^q_{\rm alg}\right)\nonumber\\
    J^g_{\omega,{\rm alg}}(\mu)&=&\frac{\alpha_s}{2\pi}\left(-\frac{1}{2}\beta_0\ln\frac{\omega^2\tan^2\frac{R}{2}}{\mu^2}+\frac{1}{2}C_A\ln^2\frac{\omega^2\tan^2\frac{R}{2}}{\mu^2}
    +d^g_{\rm alg}\right)\;,
\eea
are the unmeasured jet functions \cite{Ellis:2010rwa}, with $\beta_0=\frac{11}{3}C_A-\frac{4}{3}T_Fn_f$ and
\bea
    d^q_{\rm cone}&=&C_F\Big(\frac{7}{2}+3\ln2-\frac{5\pi^2}{12}\Big)\nonumber\\
    d^q_{\rm k_T}&=&C_F\Big(\frac{13}{2}-\frac{3\pi^2}{4}\Big)\nonumber\\
    d^g_{\rm cone}&=&C_A\Big(\frac{137}{36}+\frac{11}{3}\ln2-\frac{5\pi^2}{12}\Big)-T_Fn_f\Big(\frac{23}{18}+\frac{4}{3}\ln2\Big)\nonumber\\
    d^g_{\rm k_T}&=&C_A\Big(\frac{67}{9}-\frac{3\pi^2}{4}\Big)-T_Fn_f\Big(\frac{23}{9}\Big)\;.
\eea
Note that, the unmeasured jet functions are boost-invariant to all orders in perturbation theory, whereas the jet energy functions are boost-covariant up to terms which are power suppressed in $R$. This gives a strong constraint on the $R$ dependence of the jet and soft functions in the factorization theorem. For example, the $R$ dependence of the unmeasured jet functions can only appear as the logarithms of the ratio between $\omega\tan\frac{R}{2}$ and $\mu$. 

The differential jet shapes of quark jets at ${\cal O}(\alpha_s)$ are
\bea
    \psi^q_{{\rm cone}}(r)
    &=&\frac{\alpha_s C_F}{2\pi}\frac{1}{16\sin r}\Big[1-\Big(\frac{49+2X+X^2}{(1+X)^2}+64\ln\frac{X}{1+X}\Big)\sec^2\frac{r}{2}\Big]\nonumber\\
    \psi^q_{{k_T}}(r)
    &=&\frac{\alpha_s C_F}{2\pi}\frac{1}{\sin r}\Big(-3+6X-3X^2-4\ln X\Big)\sec^2\frac{r}{2}\;,
\eea
while for gluon jets, we have
\bea
    \psi^g_{{\rm cone}}(r)
    &=&\frac{\alpha_s}{2\pi}\frac{1}{\sin r}\left[\frac{-(11+20X+12X^2)C_A+2(2+2X+3X^2)T_Fn_f}{3(1+X)^4} \right.\nonumber\\
     &&~~~~~~~~~~\left.   -4C_A\ln\frac{X}{1+X}\sec^2\frac{r}{2}\right]\nonumber\\
    \psi^g_{{k_T}}(r)
    &=&\frac{\alpha_s}{2\pi}\frac{1}{\sin r}\left[\Big(-\frac{11}{3}+8X-6X^2+\frac{8X^3}{3}-X^4-4\ln X\Big)C_A \right.\nonumber\\
    &&~~~~~~~~~~\left. +\Big(\frac{4}{3}-4X+6X^2-\frac{16X^3}{3}+2X^4\Big)T_Fn_f\right]\;.
\eea
In the $r\ll R \ll 1$ limit, the differential jet shapes become
\bea
    \psi^q(r)&=&\frac{\alpha_s C_F}{2\pi}\frac{1}{r}\left[4\ln\frac{R}{r}-3+10({\rm or~6})\frac{r}{R}\right]\nonumber\\
    \psi^g(r)&=&\frac{\alpha_s}{2\pi}\frac{1}{r}\left[4C_A\ln\frac{R}{r}-\frac{11}{3}C_A+\frac{4}{3}T_Fn_f+(12({\rm or}~8)C_A-4T_Fn_f)\frac{r}{R}\right]\;,
\eea
for jets reconstructed using the cone (or anti-$\rm k_T$) algorithm. The difference between the differential jet shapes with different jet algorithms is of ${\cal O}(r/R)$, which can be important
at the periphery of the jet. Note that in the $r\ll R \ll 1$ limit the above results reproduce the differential jet shapes calculated at ${\cal O}(\alpha_s)$ in \cite{Seymour:1997kj}.

\begin{figure}[t]
  \begin{center}
    \includegraphics[scale=0.6, trim = 0mm 0mm 0mm 0mm , clip=true]{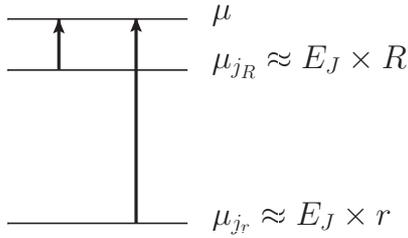}
    \caption{\label{fig:RG} The renormalization group evolution for the jet energy functions. The jet energy functions $J^{E_r}_\omega(\mu)$ of quark jets and gluon jets are calculated at ${\cal O}(\alpha_s)$ at the natural scale $\mu_{j_r}$, and they are evolved to a common renormalization scale $\mu$. At the natural scale there are no large logarithms in the jet energy functions. Large logarithms of the form $\log r/R$ in jet shapes are resummed by the renormalization group evolution between the two jet scales $\mu_{j_r}$ and $\mu_{j_R}$.}
    \end{center}
\end{figure}

\subsection{Renormalization-group equations and resummation}
\label{subsec:rg}

To resum the large logarithms we need to know how the jet energy functions evolve in energy. The renormalization-group equations satisfied by the jet energy functions are the following,
\bea
    \frac{d J^{qE_r}_\omega(\mu)}{d\ln\mu}&=&\left[-C_F\Gamma_{\rm cusp}(\alpha_s)\ln\frac{\omega^2\tan^2\frac{R}{2}}{\mu^2}-2\gamma^{q}(\alpha_s)\right]J^{qE_r}_\omega(\mu)\nonumber\\
    \frac{d J^{gE_r}_\omega(\mu)}{d\ln\mu}&=&\left[-C_A\Gamma_{\rm cusp}(\alpha_s)\ln\frac{\omega^2\tan^2\frac{R}{2}}{\mu^2}-2\gamma^{g}(\alpha_s)\right]J^{gE_r}_\omega(\mu),
\eea
where Casimir scaling is assumed to hold up to three loops and $\Gamma_{\rm cusp}$ is the cusp anomalous dimension. The anomalous dimensions $\Gamma_{\rm cusp}$ and $\gamma$ can be computed order by order in perturbation theory as series in $\frac{\alpha_s}{4\pi}$,
\bea
    \Gamma_{\rm cusp}(\alpha_s)
    &=&\Big(\frac{\alpha_s}{4\pi}\Big)\Gamma_0+\Big(\frac{\alpha_s}{4\pi}\Big)^2\Gamma_1+\cdots\;,\nonumber\\
    \gamma(\alpha_s)
    &=&\Big(\frac{\alpha_s}{4\pi}\Big)\gamma_0+\Big(\frac{\alpha_s}{4\pi}\Big)^2\gamma_1+\cdots\;.
\eea
The cusp anomalous dimension has been calculated to three loops, and $\gamma(\alpha_s)$ has only been calculated at one loop for both quark and gluon jet energy functions (and unmeasured jet functions),
\be
    \gamma^{q}_0=-3C_F\;,~\gamma^{g}_0=-\beta_0\;.
\ee
The renormalization-group equation can be solved and the jet energy function can be evolved from its natural scale $\mu_{j_r}$ to the renormalization scale $\mu$,
\be
    J^{iE_r}_\omega(\mu)=J^{iE_r}_\omega(\mu_{j_r})\exp\left[-2C_iS(\mu_{j_r},\mu)+2A_{i}(\mu_{j_r},\mu)\right]
       \left(\frac{\omega^2\tan^2\frac{R}{2}}{\mu_{j_r}^2}\right)^{C_iA_\Gamma(\mu_{j_r},\mu)}\;,
\ee
where $i=q, g$ with $C_q=C_F$ and $C_g=C_A$ the Casimir operators of the fundamental and adjoint representations in QCD. Here
\be
    S(\nu,\mu)=-\int_{\alpha_s(\nu)}^{\alpha_s(\mu)}d\alpha\frac{\Gamma_{\rm cusp}(\alpha)}{\beta(\alpha)}\int_{\alpha_s(\nu)}^\alpha\frac{d\alpha'}{\beta(\alpha')}\;,
\ee
and
\be
    A_i(\nu,\mu)=-\int_{\alpha_s(\nu)}^{\alpha_s(\mu)}d\alpha\frac{\gamma^i(\alpha)}{\beta(\alpha)}\;,~
    A_\Gamma(\nu,\mu)=-\int_{\alpha_s(\nu)}^{\alpha_s(\mu)}d\alpha\frac{\Gamma_{\rm cusp}(\alpha)}{\beta(\alpha)}\;,
\ee
are the renormalization-group evolution kernels in SCET. From these, the integral jet shape becomes
\bea
    \Psi^i_\omega(r)
    =\frac{J^{iE_r}_\omega(\mu)}{J^{iE_R}_\omega(\mu)}
   & =&\frac{J^{iE_r}_\omega(\mu_{j_r})}{J^{iE_R}_\omega(\mu_{j_R})}\exp[-2C_iS(\mu_{j_r},\mu_{j_R})+2A_{i}(\mu_{j_r},\mu_{j_R})]
    \nonumber\\
    && \times \left(\frac{\mu^2_{j_r}}{\omega^2\tan^2\frac{R}{2}}\right)^{C_iA_\Gamma(\mu_{j_R},\mu_{j_r})}\;.
\eea
Formally $\Psi_\omega(r)$ is independent of $\mu$, $\mu_{j_r}$ and $\mu_{j_R}$ to all orders in perturbation theory. However, practically we truncate the series at finite order and this induces a renormalization scale dependence. By choosing
\be
    \mu_{j_r}=\omega\tan\frac{r}{2}\approx E_J\times r\;,~~~~~\mu_{j_R}=\omega\tan\frac{R}{2}\approx E_J\times R\;,
\ee
which eliminate large logarithms in the fixed order calculations of $J^{iE_r}_\omega(\mu_{j_r})$ and $J^{iE_R}_\omega(\mu_{j_R})$ at one loop \footnote{At two loops, due to the potential issue of non-global logarithms, large logarithms in the fixed order calculations of jet energy functions may not be entirely eliminated with this scale choice. This limits the logarithmic accuracy of our resummed series to NLL.}, we can resum large logarithms of the form $\ln r/R$ in jet shapes by the renormalization group evolution of the jet energy functions between $\mu_{j_r}$ and $\mu_{j_R}$ (figure \ref{fig:RG}). The theoretical uncertainties, which come from neglecting higher order terms, can be estimated by exploiting the scale dependence in the resummed results.

\begin{figure}[t]
    \begin{center}
    \psfrag{x}{$r$}
    \psfrag{y}{$\Psi(r)$}
    \psfrag{z}{$\frac{d\Psi(r)}{dr}$}
    \includegraphics[scale=0.95, trim = 0mm 0mm 0mm 0mm , clip=true]{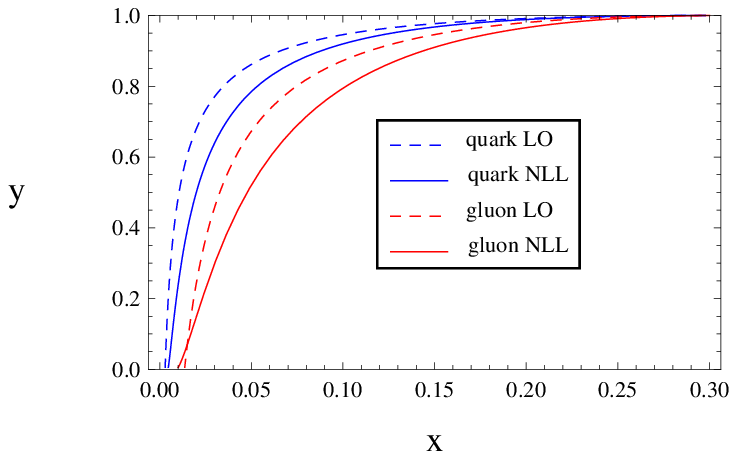}
    \includegraphics[scale=0.95, trim = 0mm 0mm 0mm 0mm , clip=true]{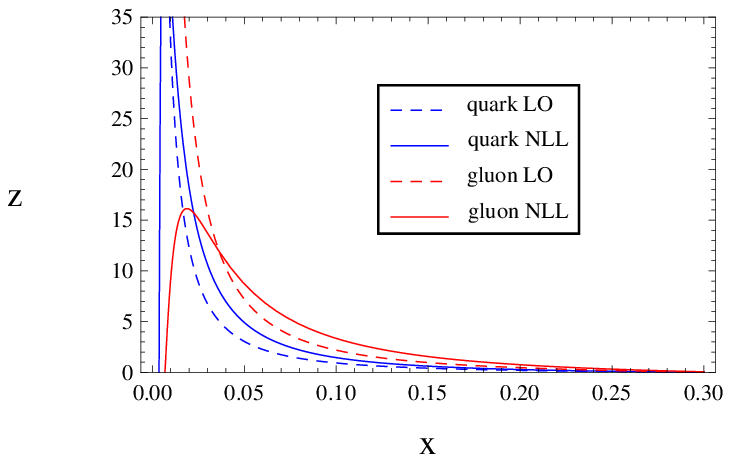}
    \caption{\label{fig:fixresum} The integral (left) and differential (right) jet shapes for {\color{blue}{quark}} and {\color{red}{gluon}} jets reconstructed using the anti-$\rm k_T$ algorithm with $R=0.3$, with a {\sl fixed} jet energy $E_J=100$ GeV plotted as an illustration. The dashed lines are the SCET calculations at leading-order (LO), whereas the solid lines are the ones at next-to-leading logarithmic order (NLL). }
    \end{center}
\end{figure}

\section{Results}
\label{sec:results}

We will compare our calculations with the {\sc pythia} 8 simulations and the CMS measurements of differential jet shapes in proton-proton collisions with nucleon-nucleon center of mass energy at $\sqrt{s_{\rm NN}}=2.76$ TeV \cite{Chatrchyan:2013kwa}. This is the reference for the studies of the jet shape modification in lead-lead collisions, which we will discuss in a forthcoming paper. Events with dijet production are the most dominant events in the experiment. Jets are reconstructed using the anti-$\rm k_T$ algorithm with $R=0.3$. This relatively small value of $R$ is chosen to reduce the background fluctuations in heavy ion collisions. The following cuts on the transverse momentum and pseudo-rapidity of a jet are imposed,
\be
    p_T^{\rm jet} > 100~{\rm GeV}\;,~0.3<|y^{\rm jet}|<2\;.
\ee
The region $|y^{\rm jet}|<0.3$ is excluded because of the techniques used in the background subtraction. Note that, the differential jet shapes measured by CMS are constructed from the transverse momenta of the charged particles with $p_T^{\rm track}>1$ GeV,
\be
    \frac{\Delta \Psi(r)}{\Delta r}=\frac{1}{N_J}\sum_{J=1}^{N_J}\frac{\Psi_J^{\rm track}(r+\delta r/2)-\Psi_J^{\rm track}(r-\delta r/2)}{\delta r}\;,
\ee
and the jet cone is divided into six annuli between $0<r<0.3$ with $\delta r=0.05$. The difference with the differential jet shapes we calculate is power suppressed by ${\cal O}(r/R)$.

\begin{figure}[t]
    \begin{center}
    \psfrag{x}{$r$}
    \psfrag{y}{$\Psi(r)$}
    \psfrag{z}{$\frac{d\Psi(r)}{dr}$}
    \includegraphics[scale=1.0, trim = 0mm 0mm 0mm 0mm , clip=true]{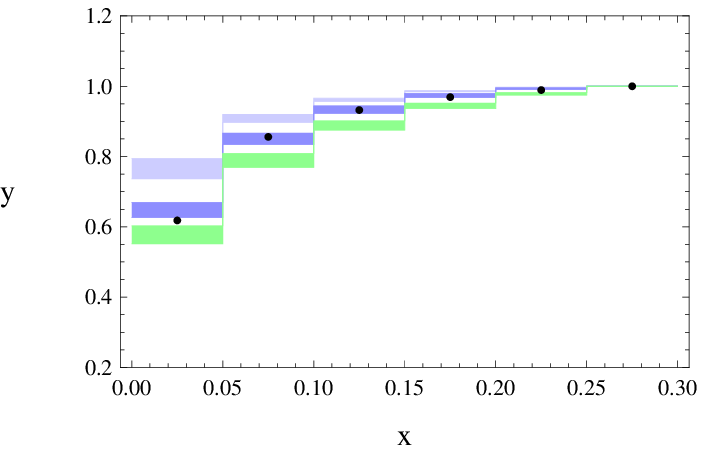}
    \includegraphics[scale=1.0, trim = 0mm 0mm 0mm 0mm , clip=true]{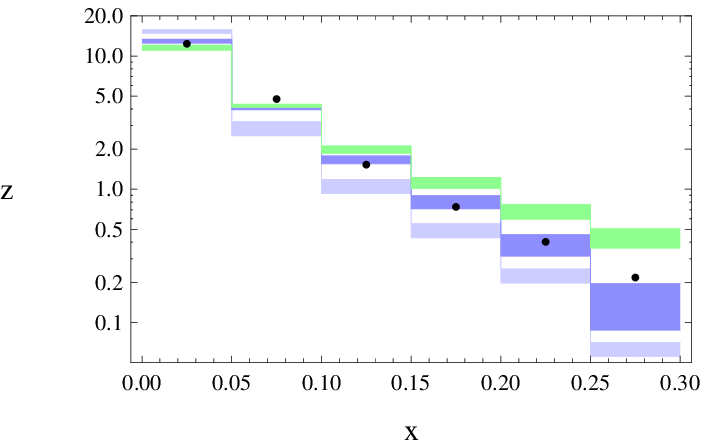}
    \caption{\label{fig:result} The integral (left) and differential (right) jet shapes in proton-proton collisions with center of mass energy at $\sqrt{s_{\rm NN}}=2.76$ TeV are plotted as a function of $r$, which is the angle from the jet axis. Jets are reconstructed using the anti-$\rm k_T$ algorithm with $R=0.3$. The cuts on the transverse momenta and rapidity of the jets ($p_T^{\rm jet}>100$ GeV and $0.3<|y^{\rm jet}|<2$) are imposed. The dots are the CMS data with negligible experimental uncertainties. The shaded blue boxes are the LO (light) and NLL (dark) results for anti-$\rm k_T$ jets, with the theoretical uncertainties estimated by varying the jet scales between $\frac{1}{2}\mu_{j_R}<\mu<2\mu_{j_R}$. As we can see, the NLL results agree with the data much better than the LO results. The shaded green boxes are the NLL results for cone jets, plotted as an illustration of the algorithm dependence in jet shapes. }
    \end{center}
\end{figure}

To look into the $p_T$ dependence of jet shape more exclusively, we also compare our calculations with the CMS differential jet shape measurement for $R=0.7$ anti-$\rm k_T$ jets in proton-proton collisions at $\sqrt{s}=7$ TeV \cite{Chatrchyan:2012mec}. Central jets with $|y^{\rm jet}|<1$ are divided into many $p_T$ bins, covering a wide range from 20 GeV to 1 TeV, and the jet cone is divided into seven annuli between $0<r<0.7$ with $\delta r=0.1$. The comparison with the first jet shape measurement at the LHC by the ATLAS collaboration \cite{Aad:2011kq} gives a similar result.

\begin{figure}[t]
    \begin{center}
    \psfrag{x}{$r$}
    \psfrag{z}{$\frac{d\Psi(r)}{dr}$}
    \includegraphics[scale=0.8, trim = 0mm 0mm 0mm 0mm , clip=true]{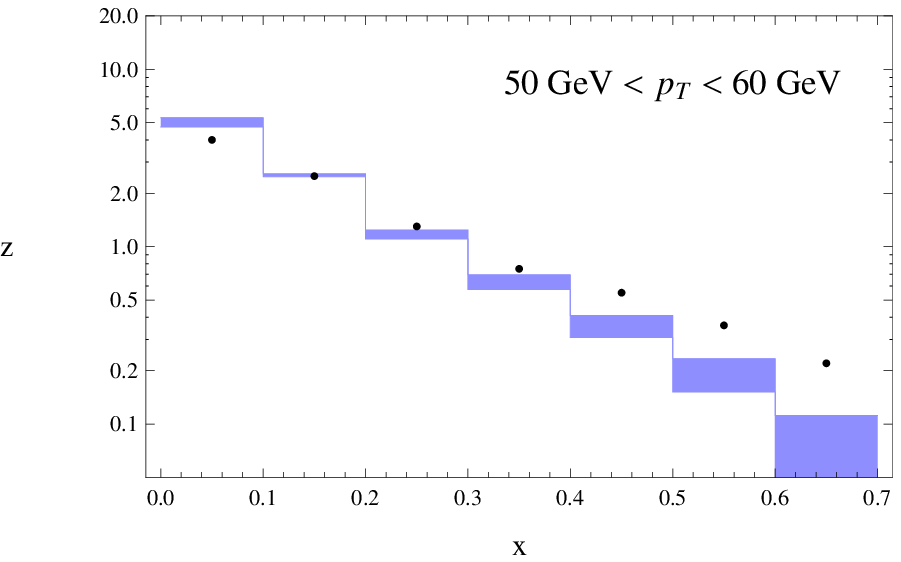}~~
    \includegraphics[scale=0.8, trim = 0mm 0mm 0mm 0mm , clip=true]{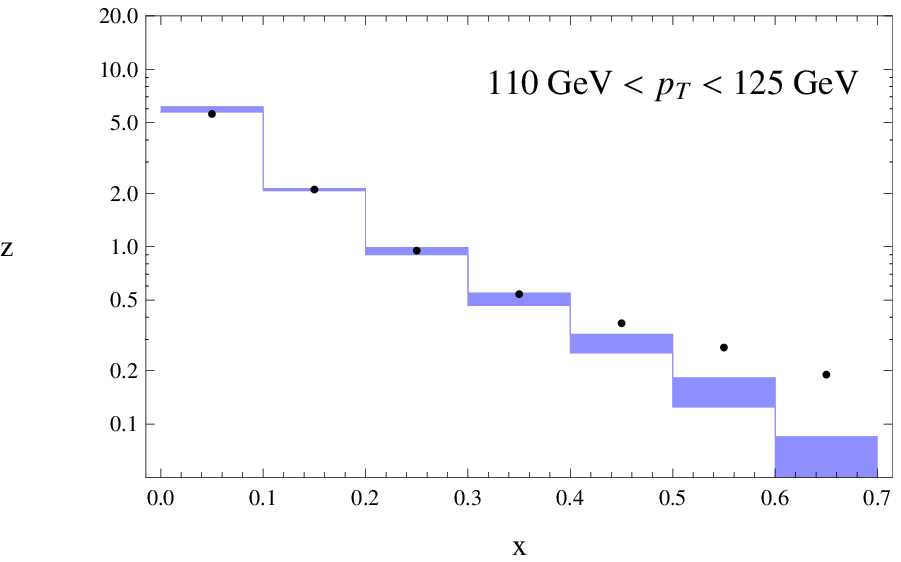}
    \includegraphics[scale=0.8, trim = 0mm 0mm 0mm 0mm , clip=true]{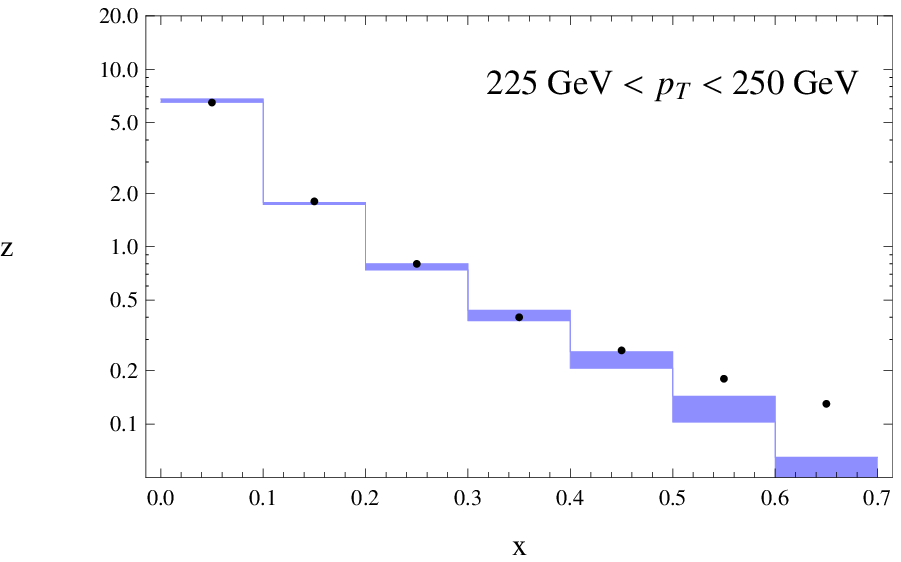}~~
    \includegraphics[scale=0.8, trim = 0mm 0mm 0mm 0mm , clip=true]{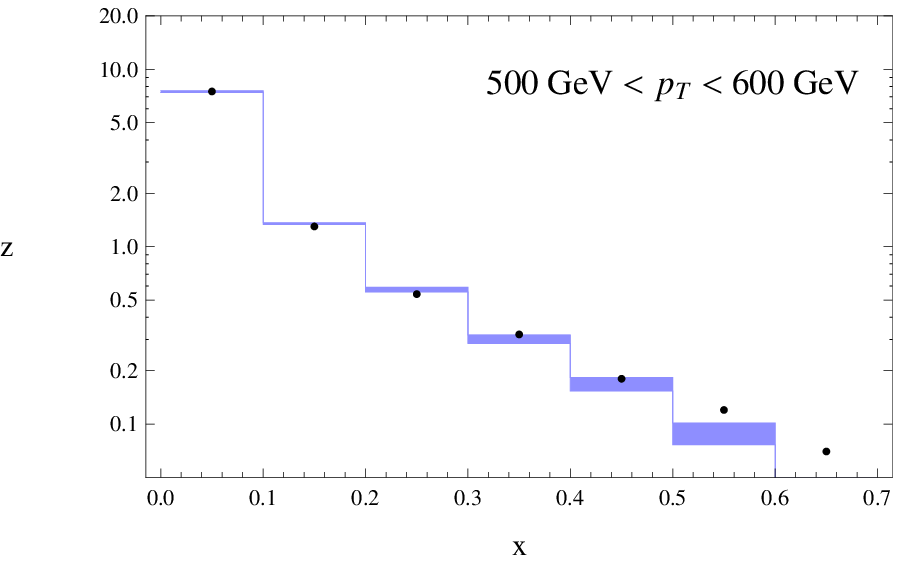}
    \caption{\label{fig:ptbin} The differential jet shapes in proton-proton collisions at $\sqrt{s}=7$ TeV. Jets are reconstructed using the anti-$\rm k_T$ algorithm with $R=0.7$, and the shapes for central jets with $|y^{\rm jet}|<1$ are examined in different $p_T$ bins: $50~ {\rm GeV} < p_T < 60 ~{\rm GeV}$, $110~ {\rm GeV} < p_T < 125~ {\rm GeV}$, $225 ~{\rm GeV} < p_T < 250~ {\rm GeV}$ and $500 ~{\rm GeV} < p_T < 600~ {\rm GeV}$, shown as examples. The dots are the CMS data with negligible experimental uncertainties. The shaded blue boxes are the NLL results, with the theoretical uncertainties estimated by varying the jet scales between $\frac{1}{2}\mu_{j_R}<\mu<2\mu_{j_R}$. For high $p_T$ jets the calculations reproduce the peak region ($r\ll R$) very well, with some discrepancy with the data in the tail region ($r\approx R$) due to power corrections. For low $p_T$ jets the power corrections become more significant.}
    \end{center}
\end{figure}

For the jet production cross section calculations, we use the CTEQ5M parton distribution functions (PDFs) \cite{Tung:2006tb} and the leading order ${\cal O}(\alpha_s^2)$ QCD results. We then average the jet shapes calculated in SCET with the fixed order QCD differential jet production cross section formula. In the SCET calculations of the differential jet shapes $\Psi_\omega(r)$, we include the one-loop jet energy functions, the two-loop cusp anomalous dimensions ($\Gamma_0$ and $\Gamma_1$) and the one-loop anomalous dimensions ($\gamma_0^{q,g}$) of the jet energy functions, as well as the two-loop running of the strong coupling constant with $\alpha_s(m_Z)=0.1172$ \cite{Becher:2008cf}. This will give us the precision formally at next-to-leading logarithmic order (NLL), including terms down to $\alpha_s^n\ln^{2n-1}\frac{r}{R}$. At next-to-next-to-leading logarithmic order (NNLL), there are potential issues related to non-global logarithms which may not be resummed using our renormalization-group technique.
Note that, unlike the non-global logarithms which arise from the existence of multiple soft scales when we consider exclusive observables sensitive to soft radiation, in this case non-global logarithms may exist within a SCET collinear sector, which is interesting. We will leave the study of non-global logarithms in jet shapes for future work \footnote{While our main focus is on the phenomenological studies of jet shapes in proton-proton and heavy ion collisions, in $e^+e^-$ collisions the full fixed order QCD calculations at NLO \cite{Catani:1996jh} and NNLO \cite{GehrmannDeRidder:2007jk,Ridder:2014wza} are available. The coefficients of the leading non-global logarithms at five loops were also recently calculated in the large $N_c$ limit \cite{Schwartz:2014wha}. They provide useful information about non-global logarithms and allow checks for precision hadronic observable resummation in $e^+e^-$ collisions beyond NLL. }.

\begin{figure}[t]
    \begin{center}
    \psfrag{x}{$r$}
    \psfrag{y}{$\Psi(r)$}
    \psfrag{z}{$\frac{d\Psi(r)}{dr}$}
    \includegraphics[scale=1.0, trim = 0mm 0mm 0mm 0mm , clip=true]{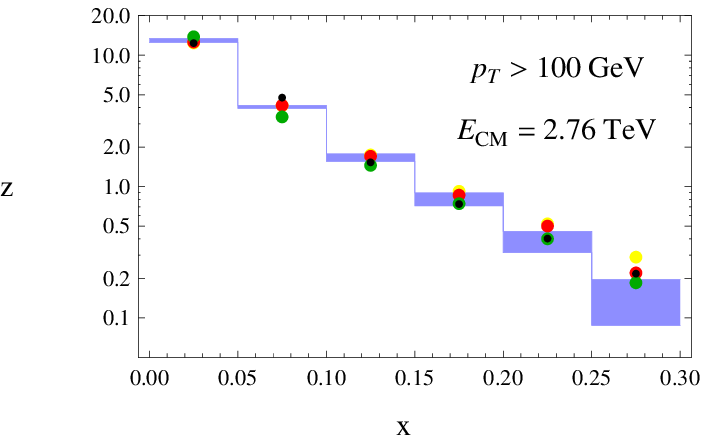}
    \includegraphics[scale=1.0, trim = 0mm 0mm 0mm 0mm , clip=true]{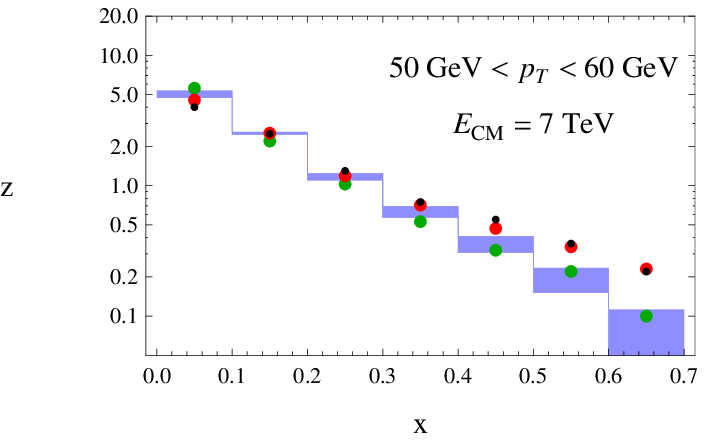}
    \caption{\label{fig:pythia} The comparison among the differential jet shapes from the CMS data, the {\sc pythia} 8 simulations, and the SCET calculations in proton-proton collisions with center of mass energy at 2.76 (left) and 7 (right) TeV. The black dots are the CMS data. The red dots are the {\sc pythia} simulation with the default tune, while the green dots are the {\sc pythia} simulation with initial state radiation (ISR) and hadronization turned off. The yellow dots in the left plot are the {\sc pythia} simulation without the $p^{\rm track}_T>$ 1 GeV cut and the background subtraction. The shaded blue boxes are the NLL SCET results, which agree with {\sc pythia} without ISR and hadronization. }
    \end{center}
\end{figure}

Figure~\ref{fig:fixresum} shows the integral and differential jet shapes of quark jets and gluon jets calculated at leading-order (LO) and next-to-leading logarithmic order (NLL) in SCET. For illustration we plot the energy distributions for jets with a fixed jet energy at 100 GeV. The fixed-order jet shape diverges at $r=0$ due to Sudakov logarithms, which need to be resummed. As we can see from the location of the peaks of the NLL differential jet shape distributions, quark jets are more localized whereas gluon jets are more spread out. Also, the effect of resummation is important throughout the whole range of $r$.

Figure \ref{fig:result} shows the comparison between our LO and NLL calculations and the CMS measurement of the integral and differential jet shapes in proton-proton collisions at $\sqrt{s_{\rm NN}}=2.76$ TeV. Jets are reconstructed using the anti-$\rm k_T$ algorithm with a small $R=0.3$. This is the reference for the studies of the jet shape modification in heavy ion collisions. The data are shown as the dots in the plot with negligible experimental uncertainties, which demands precise theoretical calculations. The shaded boxes are the theoretical uncertainties we estimate by varying the jet scales between $\frac{1}{2}\mu_{j_R}<\mu<2\mu_{j_R}$. Note that the distribution is plotted with the logarithmic scale in the vertical axis. The LO calculation, due to its divergent nature, certainly can not describe the data and resummation becomes necessary. The results for cone jets are also shown to illustrate the algorithm dependence in jet shapes. Note that cone jets are more spread out than anti-$\rm k_T$ jets.

Figure~\ref{fig:ptbin} shows a similar comparison of our calculations with the CMS differential jet shape measurement in proton-proton collisions at $\sqrt{s}=7$ TeV. Here, jets are reconstructed using the anti-$\rm k_T$ algorithm with a larger $R=0.7$, and only the central jets with $|y^{\rm jet}|<1$ are considered. We examine the differential jet shapes for jets in different $p_T$ bins: $50~ {\rm GeV} < p_T < 60 ~{\rm GeV}$, $110~ {\rm GeV} < p_T < 125~ {\rm GeV}$, $225 ~{\rm GeV} < p_T < 250~ {\rm GeV}$ and $500 ~{\rm GeV} < p_T < 600~ {\rm GeV}$, as examples. Again the dots are the CMS data with negligible experimental uncertainties, and the theoretical uncertainties are estimated by scale variation. As we can see, for high $p_T$ jets the calculations reproduce the peak region ($r\ll R$) very well, with some discrepancy with the data in the tail region ($r\approx R$) due to the power corrections of ${\cal O}(R)$. For low $p_T$ jets the power corrections of ${\cal O}(\Lambda/Q)$ become more significant because a considerable amount of radiation is outside the jets, which makes the jet more spread out. Also, this is the region where initial state radiation and non-perturbative effects also become significant.

Figure~\ref{fig:pythia} shows the comparison of the jet shapes obtained in this work with  {\sc pythia} 8 simulations at center of mass energies  2.76~TeV and 7 TeV. We turn on and off the contributions from initial state radiation (ISR) and hadronization to study their effects. The SCET calculations ignore ISR and hadronization, and they agree well with the {\sc pythia} simulation without these effects. {\sc pythia} with the default tune also reasonably agrees with the data. Note that for the jet shapes in the 2.76 TeV collisions, there are several caveats in comparing the data with the SCET resummed results. The jet shapes are reconstructed using only the charged particles with the $p^{\rm track}_T>$ 1 GeV cut. Also, to deal with the huge underlying event contamination in heavy ion collisions, an $\eta$-reflected background subtraction is performed also in reconstructing the jet shapes in proton collisions. These may affect the tail of the jet shape at about 10 to 20 $\%$ level.

\section{Summary and discussion}
\label{sec:conc}

In this paper we calculated the integral and differential jet shapes in proton-proton collisions at the LHC using soft-collinear effective theory (SCET). We performed resummation at next-to-leading logarithmic (NLL) accuracy, neglecting contributions from initial state radiation and non-perturbative effects. We aimed at obtaining a simple factorized form for the jet shape. Once we choose a recoil-free
jet axis, which always points in the collinear momentum direction, the factorized expression assumes a product form, which allows for the cancelation of the hard, unmeasured jet and soft functions in the calculation. The integral jet shape is then a ratio between two jet energy functions which we calculate at leading order (LO) for both quark jets and gluon jets reconstructed using the cone or the anti-$\rm k_T$ algorithm.

We compared our NLL calculation with the {\sc pythia} 8 simulation and the CMS measurement of jet shapes in proton-proton collisions at both~$\sqrt{s}=2.76$ TeV and~$\sqrt{s}=7$ TeV and found  good agreement. This sets the baseline calculation for the study of the jet shape modification in heavy ion collisions, which we will discuss in a forthcoming paper. We showed that the LO calculation can not describe the data well and that resummation is essential. By examining the jet shapes for jets with different transverse momenta, we found that for low $p_T$ jets the power corrections are significant. Physically, this is the region where initial-state radiation and non-perturbative effects play a role. For high $p_T$ jets the NLL resummed differential jet shape agrees with the data very well in the peak region with some room for power corrections at the very periphery of the jet.

To go beyond this precision systematically, at next-to-next-to-leading logarithmic (NNLL) accuracy we will need to calculate the two-loop jet energy function and its two-loop anomalous dimension. At this order issues about non-global logarithms and the way to resum them will also arise. The jet algorithm dependence will become more interesting because at this order we can distinguish between different recombination algorithms. It would be interesting to calculate the jet energy function, as well as the unmeasured jet function at two loops to investigate these questions. The boost properties of these jet functions can also allow us to constrain the $\log R$ dependence. On the other hand, even though the soft function  cancels in the jet shape calculation, it is of importance for the resummation of the jet rate. It would be interesting to obtain the two-loop soft function, which has been calculated for a more complicated situation of the jet thrust \cite{Kelley:2011aa,vonManteuffel:2013vja}. This simpler exercise will allow us to check the consistency of the factorization theorem of jet rate and give us insight of the possible refactorization of the soft sector without the complication of the extra measurements.



\section*{Acknowledgments}
Y.-T. C. would like to thank Andrew Hornig, Andrew Larkoski, Christopher Lee, Yen-Jie Lee, Hsiang-nan Li, Yaxian Mao, Matthew Schwartz and Wouter Waalewijn for very helpful discussions and comments on the manuscript. The authors would also like to thank the anonymous referee for careful review of the paper. Y.-T. Chien and I. Vitev are supported by the US Department of Energy, Office of Science.

\bibliographystyle{JHEP3}
\bibliography{jet_shape_FIN}

\end{document}